\documentclass[10pt,letterpaper,english]{article}
\usepackage[T1]{fontenc}
\usepackage[latin1]{inputenc}
\usepackage{amsmath}
\usepackage{graphicx}
\usepackage{amssymb}
\usepackage{array}
\makeatletter

\makeatletter
\usepackage{amscd}

\usepackage{amsfonts}

\usepackage{babel}\makeatother

\makeatother

\usepackage{babel}

\begin{document}
\newtheorem{proposition}{Proposition}[section] \newtheorem{definition}{Definition}[section]
\newtheorem{corollary}{Corollary}[section] \newtheorem{lemma}{Lemma}[section]
\newtheorem{theorem}{Theorem}[section] \newtheorem{example}{Example}[section]

\title{\textbf{Endogenous Stochastic  Arbitrage Bubbles \\
		and the Black--Scholes model.}}

\author{Mauricio Contreras G.\thanks{Universidad Metropolitana de Ciencias de la Educación UMCE.  \ \ \ \  email: mauricio.contreras@umce.cl}}

\maketitle
\noindent
This paper develops a model that incorporates the presence of stochastic arbitrage explicitly in the Black--Scholes equation. Here, the arbitrage is generated by a stochastic bubble, which generalizes the deterministic arbitrage model obtained in the literature \cite{contreras1}. It is considered to be a generic stochastic dynamic for the arbitrage bubble, and a generalized Black--Scholes equation is then derived. The resulting equation is similar to that of the stochastic volatility models, but there are no undetermined parameters as the market price of risk. \\ 
The proposed theory has asymptotic behaviors that are associated with the weak and strong arbitrage bubble limits.
For the case where the arbitrage bubble's volatility is zero (deterministic bubble), the weak limit corresponds to the usual Black-Scholes model. The strong limit case also give a Black--Scholes model, but the underlying asset's mean value replaces the interest rate. When the bubble is stochastic, the theory also has weak and strong asymptotic limits that give rise to option price dynamics that are similar to the Black--Scholes model. Explicit formulas are derived for Gaussian and lognormal stochastic bubbles. \\
Consequently, the Black--Scholes model can be considered to be a ``low energy'' limit of a more general stochastic model.
\newpage
\section{Introduction}
Since its introduction by Fischer Black, Myron Scholes \cite{BS0}, and Robert C. Merton \cite{merton}, the Black--Scholes (B--S) model has been widely used in financial engineering to price a derivative on equity. Several generalizations of the initial model premises have since been made. For example, some of these generalizations include stochastic volatility models \cite{Heston}--\cite{Gatheral}; the incorporation of jumps, which gives rise to integrodifferential equations for the option price \cite{Cont}; and, the consideration of many assets which gives its multi-asset extension \cite{Wilmott}, \cite{Bjork}, among others. \\ 
However, one of the last assumptions of the initial model to be changed was the no-arbitrage hypothesis. In effect, in the last decade, several efforts to overcome the no-arbitrage assumption have been made in the literature \cite{Otto}, \cite{Panayides}, \cite{Fedotov}. In addition, \cite{Ilinski1}, \cite{Ilinski2}, and \cite{Ilinski3} suggested that the arbitrage can be taken into account in option pricing model by changing the usual return rate of the B--S portfolio $P$ from
\begin{equation}
d P = r d t ,
\end{equation}
to 
\begin{equation} \label{arbitrage0}
d P = (r+x(t)) P d t ,
\end{equation}
where $x(t)$ follows an Ornstein--Uhlenbeck process. Using these ideas,  an endogenous arbitrage model is presented in \cite{contreras1}. Here, equation (\ref{arbitrage0}) is replaced by the stochastic differential equation
\begin{equation} \label{arbitrage_contreras}
d P = r P d t + f(S,t) P dW ,
\end{equation}
and where the deterministic function $f(S,t)$ was called an arbitrage bubble, and $dW$ is the same Brownian motion that is present in the underlying asset dynamics given by
\begin{equation} \label{dS}
d S = S \mu d t + S \sigma d W .
\end{equation}
In \cite{contreras1}, using (\ref{arbitrage_contreras}) and (\ref{dS}), the following Black--Scholes equation in the presence of an arbitrage bubble is obtained
\begin{equation} \label{BSa_equation}
\begin{aligned}
\frac{\partial V}{\partial t}+\frac{1}{2} \sigma^{2} S^{2} \frac{\partial^{2} V}{\partial S^{2}} 
+ \left(r + v(S,t) \right) \left[S \frac{\partial V}{\partial S} - V \right] = 0 & ,\\
\end{aligned}
\end{equation}
where  $V = V(S, t)$ and 
\begin{equation}
v(S,t) = \frac{\left( r - \mu \right)}{(\sigma-f(S,t))}  f(S,t) ,
\end{equation}
is a potential term that is equivalent to an electromagnetic potential that is induced by the arbitrage bubble $f(S,t)$. An approximate solution of this equation for an arbitrary bubble form $f(S,t)$ is given in \cite{contreras2} and a method  to determine the bubble $f$ from the real financial data is proposed in \cite{contreras3}. The resonances that appear in the model are also discussed in \cite{contreras4}. \\ \\
The interacting B--S equation (\ref{BSa_equation})  has two limit behaviors. The first is the ``weak bubble'' limit $f/\sigma << 1$ or $f \approx 0$, in which case the potential is $v(S,t) \approx 0$ and (\ref{BSa_equation}) becomes the usual ``free'' Black--Scholes equation 
\begin{equation} \label{BS0_equation}
\frac{\partial V}{\partial t}+\frac{1}{2} \sigma^{2} S^{2} \frac{\partial^{2} V}{\partial S^{2}} 
+ r \left[S \frac{\partial V}{\partial S} - V \right] = 0 .
\end{equation}
The second is the ``strong bubble'' limit $f/\sigma >> 1$ or $f \rightarrow \infty$, in which case
\begin{equation} 
v(S,t) = - (r-\mu) 
\end{equation}
and equation (\ref{BSa_equation}) again becomes a ``free'' Black--Scholes equation 
\begin{equation} \label{BS01_equation}
\frac{\partial V}{\partial t}+\frac{1}{2} \sigma^{2} S^{2} \frac{\partial^{2} V}{\partial S^{2}} 
+ \mu \left[S \frac{\partial V}{\partial S} - V \right] = 0 ,
\end{equation}
where the value of the interest rate has been changed to the mean of the underlying asset value $\mu$. \\ \\
In this paper, I want to incorporate possible stochastic effects on the arbitrage bubble. Hence, instead of $f$ being a given deterministic function, it becomes a random variable. I will explore its consequences on the dynamic of the option price, and I will obtain the respective weak and strong bubble limits for this case. 
\noindent
\section{The stochastic bubble}
Consider the usual underlying asset dynamics given in (\ref{dS}). Now, I generalize the deterministic bubble given in \cite{contreras1} to the stochastic case. To do that, one can assume that the arbitrage bubble satisfies the generic stochastic differential equation 
\begin{equation} \label{df}
d f =\mu_f d t+ \Gamma d W ,
\end{equation}
where $\ \mu = \mu(S, f, t) $, $\ \sigma = \sigma(S, f,t)$, $\ \mu_f = \mu_f(S, f, t) $ and $\ \Gamma = \Gamma(S, f,t) \ $ are arbitrary functions of $S$, $f$ and $t$, which defines the stochastic model completely. Note that for both equations (\ref{dS}) and (\ref{df}), there is a unique Brownian motion $dW$. Therefore, this model is endogenous in the same sense of  \cite{contreras1}. \\
In this case, the option price $V$ then also becomes a function of $f$, so $V=V(S,f,t)$ and by the It\^{o} lemma one has that
\begin{equation} \label{dV}
d V=\frac{\partial V}{\partial t} d t+\frac{\partial V}{\partial S} d S+\frac{\partial V}{\partial f} d f +\frac{1}{2} \frac{\partial^{2} V}{\partial S^{2}} d S^{2}+\frac{1}{2} \frac{\partial^{2} V}{\partial f^{2}} d f^{2}+\frac{\partial^{2} V}{\partial S \partial f} d S df ,
\end{equation}
and by replacing (\ref{dS}) and (\ref{df}) in (\ref{dV}), one has that
\begin{equation} \label{dV2}
\begin{aligned}
d V=&\ L(V) \ dt +\left[ \sigma S \frac{\partial V}{\partial S}  + \Gamma \frac{\partial V}{\partial f}   \right] d W ,
\end{aligned}
\end{equation}
where $L(V)$ denotes the differential operator action
\begin{equation} \label{L(V)}
L(V) = \frac{\partial V}{\partial t}+\frac{1}{2} \sigma^{2} S^{2} \frac{\partial V}{\partial S^{2}}+\frac{1}{2} \Gamma^{2} \frac{ \partial^{2} V}{\partial f^{2}} + S \sigma \Gamma \frac{\partial^{2} V}{\partial S \partial f}  + S \mu_{S} \frac{\partial V}{\partial S} + \mu_{f} \frac{\partial V}{\partial f} .
\end{equation}
To derive the corresponding Black--Scholes equation, one must consider a portfolio $P$ that is constructed by a number of $N_V$ options and $N_S$ underlying assets according to
\begin{equation} \label{P}
P=N_{S} S+N_{V} V ,
\end{equation}
so one has that (see \cite{Bjork}, \cite{Wilmott})
\begin{equation}
d P=N_{S} d S+N_{V} d V .
\end{equation}
According to equation (\ref{arbitrage_contreras}) \cite{contreras1}, the portfolio return in the presence of a arbitrage bubble $f$ has the form
\begin{equation}
d P= P r d t + P f d W ,
\end{equation}
so
\begin{equation} \label{dP=dP}
N_{S} d S+N_{V} d V = P r d t + P f d W .
\end{equation}
By replacing (\ref{dS}), (\ref{dV2}) in (\ref{dP=dP}), one obtains
\begin{equation}
N_{S}\left(\mu S d t + \sigma S dW \right) + N_{V} \left(L d t + \sigma S \frac{\partial V}{\partial S} d W + \Gamma \frac{ \partial V}{\partial f } d W \right) = \left(N_{s} S+N_{V} V\right) r d t + \left(N_{s} S+N_{V} V\right) f d t . 
\end{equation}  
By equalling terms in $dt$ and $dW$ in this equation, one finds the system
\begin{equation} \label{linearsystem1}
\begin{array}{l}
\left(\mu_{S} S- S r\right) N_{S} + (L- r V) N_{V}=0 \\
(\sigma S - S f) N_{S} + \left(\sigma S \frac{\partial V}{\partial S}+ \Gamma \frac{\partial V}{\partial f} - V f \right) N_{V} = 0 ,
\end{array}
\end{equation}
To obtain a solution with $N_S \ne 0$ and $N_V \ne 0$, the determinant associated to the matrix form of this system (\ref{linearsystem1}) must be equal to zero; that is,
\begin{equation}
\left(\mu S- S r\right)\left(\sigma S \frac{\partial V}{\partial S}+ \Gamma \frac{\partial V}{\partial f} - V f\right)-(\sigma S - S  f)(L - r V) = 0 ,
\end{equation} 
that is,
\begin{equation}
(L-r V)=\frac{\left(\mu S - S r \right) \left(\sigma S \frac{\partial V}{\partial S}+ \Gamma \frac{\partial V}{\partial f} - V f\right)}{(\sigma S - S f)} .
\end{equation} 
Now, by replacing $L(V)$ in (\ref{L(V)}) and simplifying terms, one finally ends with the following explicit Black--Scholes equation for the option price
\begin{equation} \label{BSf_equation}
\begin{aligned}
\frac{\partial V}{\partial t}+\frac{1}{2} \sigma^{2} S^{2} \frac{\partial^{2} V}{\partial S^{2}}+\frac{1}{2} \Gamma^{2} \frac{\partial^{2} V}{\partial f^{2}}+S \sigma \Gamma \frac{\partial^{2} V}{\partial S \partial f} + \left(r + v(f) \right) \left[S \frac{\partial V}{\partial S} - V \right] + \left(\mu_{f}-\frac{\left(\mu -r\right)}{(\sigma-f)} \Gamma \right) \frac{\partial V}{\partial f} = 0. & \\
\end{aligned}
\end{equation}
where  \\
\begin{equation}
v(f) = \frac{\left( r - \mu \right)}{(\sigma-f)}  f ,
\end{equation}
is the  ``electromagnetic'' potential mentioned in \cite{contreras1}. Note that the (\ref{BSf_equation}) is the same form of the Blacks--Scholes equation for a stochastic volatility model, but without external undetermined functions as the market price of risk \cite{Wilmott},  \cite{Gatheral}. \\ \\
For the $ \Gamma = 0 $ case, equation (\ref{BSf_equation}) reduces to
\begin{equation} \label{BSf_equation_gammacero}
\begin{aligned}
\frac{\partial V(S,f,t)}{\partial t}+\frac{1}{2} \sigma^{2} S^{2} \frac{\partial^{2} V(S,f,t)}{\partial S^{2}}+ \left(r + v(f) \right) \left[S \frac{\partial V(S,f,t)}{ \partial S} - V(S,f,t) \right] + \mu_{f} \frac{\partial V(S,f,t)}{\partial f} = 0. & \\
\end{aligned}
\end{equation}
Here, $f$ is, due to (\ref{df}), the deterministic function 
\begin{equation} 
\frac{df}{dt} =\mu_f(S, f, t) ,
\end{equation}
so $f = f(S,t)$ and the option price becomes a function of $S$ and $t$ only, which is defined by
\begin{equation} 
V(S,t) = V(S,f(S,t),t) ,
\end{equation}
This means that
\begin{equation} 
\frac{\partial V(S,t)}{\partial t} = \frac{\partial V(S,f(S,t),t)}{\partial f} \frac{ d f(S,t)}{d t} + \frac{\partial V(S,f(S,t),t)}{\partial t} ,
\end{equation}
that is,
\begin{equation} 
\frac{\partial V(S,t)}{\partial t} = \frac{\partial V(S,f(S,t),t)}{\partial f} \ \mu_f + \frac{\partial V(S,f(S,t),t)}{\partial t} ,
\end{equation}
Consequently, in terms of $V(S,t)$, equation (\ref{BSf_equation_gammacero}) is finally
\begin{equation} 
\begin{aligned}
\frac{\partial V(S, t)}{\partial t}+\frac{1}{2} \sigma^{2} S^{2} \frac{\partial^{2} V(S, t)}{\partial S^{2}}+ \left(r + v(f) \right) \left[S \frac{\partial V(S, t)}{ \partial S} - V(S, t) \right] = 0. & \\
\end{aligned}
\end{equation}
which is the same equation (\ref{BSa_equation}). Thus, the case $ \Gamma = 0 $ recovers the deterministic arbitrage bubble case. \\ \\
In the rest of this paper, I will test the effect of the stochastic bubble on the Black--Scholes solution on analytical grounds.  I will analyze two special cases: one is the Gaussian bubble, and the other is the lognormal bubble. For these two models, one can find an analytical solution valid for some asymptotic regions in the $(S, f, t)$ space. \\ 
Of course, for more general models, to find solutions of equation (\ref{BSf_equation}) one must use numerical methods \cite{Wilmott}. Nevertheless, the analytical solutions obtained in this work can be used to test the grade of exactitude of the numerical solutions. In a further incoming paper, I will tackle the numerical analysis in a detailed manner and I will then compare it with the analytical solutions obtained in the following sections. \\

\section{The Gaussian bubble}
For the Gaussian bubble, one can consider that the asset's dynamics (\ref{dS}) is given by the usual Black--Scholes case; that is, $\mu$ and $\sigma$ are constants. In addition, for the Gaussian bubble, the $f$--dynamic is given by (\ref{df}) with $\mu_f$ and $\Gamma$ constants. In fact, these parameters represent the mean height and the variance of the bubble. \\
Thus, one needs to find solutions of (\ref{BSf_equation}), with all parameters being constant. An analytical solution can be obtained that is valid in the following regions of the $(S, f, t)$ space: \\ \\
{\bf (a)} the region $f \approx 0$, in which case $v(f) = \frac{\left( r - \mu \right)}{(\sigma-f)}  f \approx 0$ and (\ref{BSf_equation}) takes the form
\begin{equation} \label{BSf_equation_f=0}
\begin{aligned}
\frac{\partial V}{\partial t}+\frac{1}{2} \sigma^{2} S^{2} \frac{\partial^{2} V}{\partial S^{2}}+\frac{1}{2} \Gamma^{2} \frac{\partial^{2} V}{\partial f^{2}}+S \sigma \Gamma \frac{\partial^{2} V}{\partial S \partial f} + r \left[S \frac{\partial V}{\partial S} - V \right] + \left(\mu_{f}-\frac{\left(\mu -r\right)}{\sigma} \Gamma \right) \frac{\partial V}{\partial f} = 0 , & \\
\end{aligned}
\end{equation}
and \\ \\
{\bf (b)} the asymptotic limit $\ f >> \sigma $ \ or \ $f \rightarrow \infty$, in which case 
$v(f) = \frac{\left( r - \mu \right)}{(\sigma-f)}  f  \rightarrow -(r-\mu) $ so the  asymptotic Black--Scholes equation becomes
\begin{equation} \label{BSf_equation_f=infinity}
\begin{aligned}
\frac{\partial V}{\partial t}+\frac{1}{2} \sigma^{2} S^{2} \frac{\partial^{2} V}{\partial S^{2}}+\frac{1}{2} \Gamma^{2} \frac{\partial^{2} V}{\partial f^{2}}+S \sigma \Gamma \frac{\partial^{2} V}{\partial S \partial f} + \mu \left[S \frac{\partial V}{\partial S} - V \right] + \mu_{f} \frac{\partial V}{\partial f} = 0 .& \\
\end{aligned}
\end{equation}
One can consider equation (\ref{BSf_equation_f=0}) as the  ``weak bubble limit'' of (\ref{BSf_equation}), whereas (\ref{BSf_equation_f=infinity}) can be considered as the ``strong bubble limit'' of (\ref{BSf_equation}). \\ \\
Instead of working directly on equations (\ref{BSf_equation_f=0}) and (\ref{BSf_equation_f=infinity}) to obtain the analytical solutions, one can again consider the ``full'' equation (\ref{BSf_equation}) for the Gaussian bubble, and take the following transformation
\begin{equation}
\left\{\begin{array}{l}
\bar{u}= \ln S - \left(r-\frac{1}{2} \sigma^{2}\right) t \\
f=f \\
t=t .
\end{array}\right.
\end{equation}
This maps (\ref{BSf_equation})  to the following equation
\begin{equation}
\begin{aligned}
\frac{\partial V}{\partial t}+\frac{1}{2} \sigma^{2} \frac{\partial V}{\partial \bar{u}^{2}}+\frac{1}{2} \Gamma^2 \frac{\partial^{2} V}{\partial f^{2}}+\sigma \Gamma \frac{\partial^{2} V}{\partial \bar{u} \partial f}+v(f) \frac{\partial V}{\partial \bar{u}} \\
+\left(\mu_{f}-\frac{(\mu-r) \Gamma}{(\sigma-f)}\right) \frac{\partial V}{\partial f}-(r+v(f)) V=0 .
\end{aligned}
\end{equation}
By defining 
\begin{equation}  \label{Vpsi_gauss}
V(\bar{u},f,t) = e^{-r(T-t)} \psi(\bar{u},f,t)
\end{equation}
one has that 
\begin{equation}
\frac{\partial \psi}{\partial t}+\left(\frac{1}{2} \sigma^{2} \frac{\partial^{2} \psi}{\partial \bar{u}^{2}}+\frac{1}{2} \Gamma^{2} \frac{\partial^{2} \psi}{\partial f^{2}}+\sigma \Gamma \frac{\partial^{2} \psi}{\partial \bar{u} \partial f}\right)+ \\
v(f)\left(\frac{\partial \psi}{\partial \bar{u}}-\psi\right) +\left(\mu_{f}-\frac{(\mu-r) \Gamma}{(\sigma-f)}\right) \frac{\partial \psi}{\partial f}=0 .
\end{equation}
Now, by performing the following transformation
\begin{equation}
\left\{ \begin{array}{l}
\bar{x}=\frac{1}{2}\left(\frac{\bar{u}}{\sigma}+\frac{f}{\Gamma}\right) - \frac{\mu_{f}}{2 \Gamma} t\\
\bar{y}=\frac{1}{2}\left(\frac{\bar{u}}{\sigma}-\frac{f}{\Gamma}\right) + \frac{\mu_{f}}{2 \Gamma} t \\
\tau = T-t ,
\end{array} \right.
\end{equation}
one arrives to
\begin{equation}
-\frac{\partial \psi}{\partial \tau}+\frac{1}{2} \frac{\partial^{2} \psi}{\partial \bar{x}^{2}}+\left[\frac{1}{2 \sigma} v(f)-\frac{1}{2} \frac{(\mu-r)}{(\sigma-f)}\right] \frac{\partial \psi}{\partial \bar{x}}+\left[\frac{1}{2 \sigma} v(f)+\frac{1}{2} \frac{(\mu-r)}{(\sigma-f)}\right] \frac{\partial \psi}{\partial \bar{y}} -v(f) \psi=0 ,
\end{equation}
where $f$ denotes the function
\begin{equation}
f = f(\bar{x}, \bar{y}, \tau ) = \Gamma \left(\bar{x} - \bar{y} - \frac{\mu_{f}}{\Gamma} (T-\tau) \right) .
\end{equation}
Now, by replacing $v(f)$ explicitly one has that
\begin{equation} \label{BSE_xbar_ybar_gauss}
-\frac{\partial \psi}{\partial \tau}+\frac{1}{2} \frac{\partial^{2} \psi}{\partial \bar{x}^{2}}+
\left[\frac{(r - \mu)}{2 \sigma} \frac{(1+f/\sigma)}{(1-f/\sigma)}\right] \frac{\partial \psi}{\partial \bar{x}}+
\left[ -\frac{(r - \mu)}{2 \sigma} \right] \frac{\partial \psi}{\partial \bar{y}} - \frac{\left( r - \mu \right)   }{(1-f/\sigma)} (f/\sigma)  \ \psi=0 ,
\end{equation}
One can now consider the ``weak'' and ``strong'' limits of (\ref{BSE_xbar_ybar_gauss}). 
\subsection{The weak bubble limit for the Gaussian bubble}
The ``weak'' bubble limit, that is
\begin{equation}
f/\sigma << 1 ,
\end{equation}
will be valid in the  $(\bar{x}, \bar{y}, \tau )$ space  region for which 
\begin{equation}
 \Gamma \left(\bar{x} - \bar{y} - \frac{\mu_{f}}{\Gamma} (T-\tau) \right) << \sigma ,
\end{equation}
and the equation (\ref{BSE_xbar_ybar_gauss}) can be approximated to ($ f \approx 0$) 
\begin{equation} \label{BSE_xbar_ybar_f=0_gauss}
-\frac{\partial \psi}{\partial \tau}+\frac{1}{2} \frac{\partial^{2} \psi}{\partial \bar{x}^{2}}+
\left[\frac{(r - \mu)}{2 \sigma} \right] \frac{\partial \psi}{\partial \bar{x}} - 
\left[\frac{(r - \mu)}{2 \sigma} \right] \frac{\partial \psi}{\partial \bar{y}} =0 .
\end{equation}
Note that this equation also can be obtained by doing  several coordinate transformations directly to equation (\ref{BSf_equation_f=0}).
\subsection{The strong bubble limit for the Gaussian bubble}
The ``strong'' bubble limit, that is
\begin{equation}
f/\sigma >> 1 ,
\end{equation}
will be valid in the  $(\bar{x}, \bar{y}, \tau )$ space  region for which 
\begin{equation}
\Gamma \left(\bar{x} - \bar{y} - \frac{\mu_{f}}{\Gamma} (T-\tau) \right) >> \sigma ,
\end{equation}
so equation (\ref{BSE_xbar_ybar_gauss}) can be approximated in this region with the limit $f \rightarrow \infty$, so 
\begin{equation}
\frac{(1+f/\sigma)}{(1-f/\sigma)} \rightarrow -1 ,
\end{equation}
and
\begin{equation}
v(f) \rightarrow -(r-\mu) ,
\end{equation}
so asymptotic Black--Scholes equation is
\begin{equation} 
-\frac{\partial \psi}{\partial \tau}+\frac{1}{2} \frac{\partial^{2} \psi}{\partial \bar{x}^{2}} -
\left[\frac{(r - \mu)}{2 \sigma} \right] \frac{\partial \psi}{\partial \bar{x}} -
\left[\frac{(r - \mu)}{2 \sigma} \right] \frac{\partial \psi}{\partial \bar{y}} + \left( r - \mu \right) \ \psi=0 .
\end{equation}
Now by defining 
\begin{equation}
\psi = e^{(r-\mu) \tau} \Psi
\end{equation}
one can find the strong limit in terms of $\Psi$ as
\begin{equation} \label{BSE_xbar_ybar_f=infinito_gauss}
-\frac{\partial \Psi}{\partial t}+\frac{1}{2} \frac{\partial^{2} \Psi}{\partial \bar{x}^{2}}-\left[\frac{(r-\mu)}{2 \sigma}\right] \frac{\partial \Psi}{\partial \bar{x}}-\left[\frac{(r-\mu)}{2 \sigma}\right] \frac{\partial \Psi}{\partial \bar{y}}=0 .
\end{equation}
Note that equation (\ref{BSE_xbar_ybar_f=infinito_gauss}) has the same form of equation (\ref{BSE_xbar_ybar_f=0_gauss}) for $\psi$ and  note also that the option price in the strong limit  is given by 
\begin{equation} \label{Vpsi_gauss_2}
V = e^{-r(T-t)} \psi =  e^{-r(T-t)}  e^{(r-\mu)(T-t)} \Psi = e^{-\mu (T-t)} \Psi .
\end{equation}
\subsection{The case $f/\sigma \approx -1$ for the Gaussian bubble}
For the Gaussian case, the variable $f$ can take negative values; thus, $f$ could also take values near $-\sigma$. Then, when $f/\sigma \approx -1$, that is for
region in the $(\bar{x}, \bar{y}, \tau)$ for which 
\begin{equation}
\Gamma \left(\bar{x} - \bar{y} - \frac{\mu_{f}}{\Gamma} (T-\tau) \right) \approx - \sigma ,
\end{equation} 
equation (\ref{BSE_xbar_ybar_gauss}) reduces to
\begin{equation} 
-\frac{\partial \psi}{\partial \tau}+\frac{1}{2} \frac{\partial^{2} \psi}{\partial \bar{x}^{2}}+
\left[ -\frac{(r - \mu)}{2 \sigma} \right] \frac{\partial \psi}{\partial \bar{y}} + \frac{\left( r - \mu \right)   }{2}  \ \psi=0 .
\end{equation}
By defining
\begin{equation}
\psi = e^{\frac{(r-\mu)}{2} \tau} \Psi ,
\end{equation}
so
\begin{equation} \label{BSE_xbar_ybar_gauss_f=-1}
-\frac{\partial \Psi}{\partial \tau}+\frac{1}{2} \frac{\partial^{2} \Psi}{\partial \bar{x}^{2}}-\frac{(r-\mu)}{2 \sigma} \frac{\partial^{2} \Psi}{\partial \bar{y}^{2}}=0 .
\end{equation}
Note that the option price in this case is
\begin{equation}  \label{Vpsi_gauss_f=-1}
V= e^{-r \tau} \psi = e^{-r \tau} e^{\frac{(r-\mu)}{2} \tau} \Psi = e^{\frac{-(r+\mu)}{2} \tau} \Psi .
\end{equation}

\section{The lognormal bubble}
For the  lognormal bubble, the underlying $S$--dynamics are the same as the Gaussian but for $f$ one takes instead
\begin{equation} 
\begin{array}{l}
\mu_f = f \ \bar{\mu}_f \\
\Gamma = f \ \bar{\Gamma} ,
\end{array}
\end{equation}
where $\bar{\mu}_f$ and $\bar{\Gamma}$ are constants. Consequently, (\ref{df}) becomes
\begin{equation}
d f =\bar{\mu}_{f} f d t+f \bar{\Gamma} d W .
\end{equation}
In this case, both the underlying asset and the stochastic bubble have lognormal dynamics. For this case, equation (\ref{BSf_equation}) becomes
\begin{equation} \label{BSf_equation_lognormal}
\begin{aligned}
\frac{\partial V}{\partial t}+\frac{1}{2} \sigma^{2} S^{2} \frac{\partial^{2} V}{\partial S^{2}}+\frac{1}{2} \bar{\Gamma}^{2} f^2  \frac{\partial^{2} V}{\partial f^{2}}+  \sigma \bar{\Gamma} S f \frac{\partial^{2} V}{\partial S \partial f} + \left(r + v(f) \right) \left[S \frac{\partial V}{\partial S} - V \right] + \left(\bar{\mu}_{f}-\frac{\left(\mu -r\right)}{(\sigma-f)} \bar{\Gamma} \right) f \frac{\partial V}{\partial f} = 0 , & \\
\end{aligned}
\end{equation}
Now by taking the coordinate transformation
\begin{equation} \label{transformationubar=u_vbar=v_lognormal}
\left\{ \begin{array}{ll}
\bar{u} &=\ln S - \left(r-\frac{1}{2} \sigma^{2}\right) t \\
\bar{v} &=\ln f - \left(\bar{\mu}_{f}-\frac{1}{2} \bar{\Gamma}^{2}\right) t \\
t &=t ,
\end{array} \right.
\end{equation}
and defining 
\begin{equation} 
V(\bar{u}, \bar{v}, t) = e^{-r(T-t)} \psi (\bar{u}, \bar{v}, t) ,
\end{equation}
equation (\ref{BSf_equation_lognormal}) maps to
\begin{equation} \label{BSf_equation_lognormal_ubar_vbar}
\begin{aligned}
\frac{\partial \psi}{\partial t}+\frac{1}{2} \sigma \frac{\partial^{2} \psi}{\partial \bar{u}^{2}}+\frac{1}{2} \bar{\Gamma}^{2} \frac{\partial^{2} \psi}{\partial \bar{v}^{2}}+\sigma \Gamma \frac{\partial^{2} \psi}{\partial \bar{u} \partial \bar{v}} \\
+v(f)\left(\frac{\partial \psi}{\partial \bar{u}}-\psi\right)-\frac{(\mu-r)}{(\sigma-f)} \bar{\Gamma} \frac{\partial \psi}{\partial \bar{v}}=0 .
\end{aligned}
\end{equation}
Now, by doing the following transformation
\begin{equation} 
\left\{ \begin{array}{ll}
x=\frac{1}{2}\left(\frac{\bar{u}}{\sigma}+\frac{\bar{v}}{\bar{\Gamma}}\right) &   \\ \\
y=\frac{1}{2} \left(\frac{\bar{u}}{\sigma}-\frac{\bar{v}}{\bar{\Gamma}}\right) &  \\ \\
\tau = T-t ,
\end{array} \right.
\end{equation}
the equation (\ref{BSf_equation_lognormal_ubar_vbar}) gets
\begin{equation}
- \frac{\partial \psi}{\partial \tau}+\frac{1}{2} \frac{\partial^{2} \psi}{\partial x^{2}}+\left(\frac{1}{2 \sigma} v(f)-\frac{1}{2} \frac{(\mu-r)}{(\sigma-f)}\right) \frac{\partial \psi}{\partial x} \\
+\left(\frac{1}{2 \sigma} v(f)+\frac{1}{2} \frac{(\mu-r)}{(\sigma-f)}\right) \frac{\partial \psi}{\partial y}=0.
\end{equation}
where $f$ denotes the function
\begin{equation} \label{fpositive_lognornal}
f=f(x,y, \tau)=e^{\bar{\Gamma}(x-y)+\left(\bar{\mu}_{f}-\frac{1}{2} \bar{\Gamma}^{2}\right) (T-\tau)} .
\end{equation}
By replacing $v(f)$, one finally obtains 
\begin{equation}  \label{BSf_equation_lognormal_x_y_tau}
- \frac{\partial \psi}{\partial \tau}+\frac{1}{2} \frac{\partial^{2} \psi}{\partial x^{2}}+\left[\frac{(r-\mu)}{2 \sigma} \frac{(1+f / \sigma)}{(1-f / \sigma)}\right] \frac{\partial \psi}{\partial x}-\frac{(r-\mu)}{2 \sigma} \frac{\partial \psi}{\partial y}=0.
\end{equation}

\subsection{The weak bubble limit for the lognormal bubble}
Note that  when $ f/ \sigma << 1 $, that is in the time-spatial $(x,y, \tau)$ region, that
\begin{equation}
\bar{\Gamma}(x-y)+\left(\bar{\mu}_{f}-\frac{1}{2} \bar{\Gamma}^{2}\right) (T-\tau)  << \ln \sigma ,
\end{equation}
then
\begin{equation}
\frac{(1+f / \sigma)}{(1-f / \sigma)} \approx  1 ,
\end{equation}
so the Black--Scholes equation (\ref{BSf_equation_lognormal_x_y_tau}) can be approximated in this region by
\begin{equation} \label{BSE_xbar_ybar_f=0_lognormal}
- \frac{\partial \psi}{\partial \tau}+\frac{1}{2} \frac{\partial^{2} \psi}{\partial x^{2}}+\frac{(r-\mu)}{2 \sigma} \frac{\partial \psi}{\partial x}-\frac{(r-\mu)}{2 \sigma} \frac{\partial \psi}{\partial y}=0 .
\end{equation}
\subsection{The strong bubble limit for the lognormal bubble}
For the case $ f/ \sigma >> 1 $, that is in the time-spatial $(x,y, \tau)$ region, that
\begin{equation}
\bar{\Gamma}(x-y)+\left(\bar{\mu}_{f}-\frac{1}{2} \bar{\Gamma}^{2}\right) (T-\tau)  >> \ln \sigma ,
\end{equation}
then
\begin{equation}
\frac{(1+f / \sigma)}{(1-f / \sigma)} \approx  - 1 ,
\end{equation}
so the Black--Scholes equation (\ref{BSf_equation_lognormal_x_y_tau}) gets to the asymptotic equation
\begin{equation} \label{BSE_xbar_ybar_f=infinito_lognormal}
- \frac{\partial \psi}{\partial \tau}+\frac{1}{2} \frac{\partial^{2} \psi}{\partial x^{2}}-\frac{(r-\mu)}{2 \sigma} \frac{\partial \psi}{\partial x}-\frac{(r-\mu)}{2 \sigma} \frac{\partial \psi}{\partial y}=0 .
\end{equation}
Note that due to (\ref{fpositive_lognornal}), $f$ can take only positive values. Therefore, there is no analog to $\ f/\sigma = -1 \ $ case for the lognormal bubble.
\section{The analytical solutions}
The asymptotic equations (\ref{BSE_xbar_ybar_f=0_gauss}), (\ref{BSE_xbar_ybar_f=infinito_gauss}), (\ref{BSE_xbar_ybar_gauss_f=-1}),  (\ref{BSE_xbar_ybar_f=0_lognormal}) and (\ref{BSE_xbar_ybar_f=infinito_lognormal}) are particular cases of the generic equation
\begin{equation} \label{BSEffreegeneralcase}
-\frac{\partial \psi}{\partial \tau}+\frac{1}{2} \frac{\partial^{2} \psi}{\partial x^{2}}+\alpha_{x} \frac{\partial \psi}{\partial x}+\alpha_{y} \frac{\partial \psi}{\partial y}=0 ,
\end{equation}
where $\alpha_{x}$ and $\alpha_{y}$ are constants. In fact, the propagator of (\ref{BSEffreegeneralcase}) is
\begin{equation} \label{Propagatorfreegeneralcase}
P(x, y, \tau)=\frac{1}{\sqrt{2 \pi \tau}} \ e^{-\frac{(x + \alpha_x \tau )^{2}}{2 \tau}} \delta \left(y + \alpha_{y} \tau\right) ,
\end{equation}
where $\delta(x)$ is the Dirac's delta function. So, if $\Phi(x,y)$ is some initial condition for equation (\ref{BSEffreegeneralcase}), then its solution is 
\begin{equation}
\psi(x,y, \tau) = \int_{-\infty}^{+\infty}  \int_{-\infty}^{+\infty}  P(x-x', y-y', \tau) \ \Phi(x',y') \ dx' dy'  ,
\end{equation}
that is
\begin{equation} \label{solution_generic}
\psi(x,y, \tau) = \int_{-\infty}^{+\infty}  \int_{-\infty}^{+\infty}  \frac{1}{\sqrt{2 \pi \tau}} \ e^{-\frac{(x-x' + \alpha_x \tau )^{2}}{2 \tau}} \delta \left(y-y' + \alpha_{y} \tau\right) \ \Phi(x',y') \ dx' dy' .
\end{equation}

\subsection{The solutions for the Gaussian bubble}
The weak and  strong limits of the Gaussian bubble are given by equations (\ref{BSE_xbar_ybar_f=0_gauss}), (\ref{BSE_xbar_ybar_f=infinito_gauss}), which generically can be written as
\begin{equation} \label{solution_generic_xbar_ybar}
-\frac{\partial \psi}{\partial \tau}+\frac{1}{2} \frac{\partial^{2} \psi}{\partial \bar{x}^{2}}+\alpha_{\bar{x}} \frac{\partial \psi}{\partial \bar{x}}+\alpha_{\bar{y}} \frac{\partial \psi}{\partial \bar{y}}=0 .
\end{equation}
The solution  (\ref{solution_generic}) is then given by
\begin{equation} \label{solution_generic_gauss}
\psi(\bar{x},\bar{y}, \tau) = \int_{-\infty}^{+\infty}  \int_{-\infty}^{+\infty}  \frac{1}{\sqrt{2 \pi \tau}} \ e^{-\frac{(\bar{x}-\bar{x}' + \alpha_{\bar{x}} \tau )^{2}}{2 \tau}} \delta \left(\bar{y}-\bar{y}' + \alpha_{\bar{y}} \tau \right) \ \Phi(\bar{x}',\bar{y}') \ d\bar{x}' d\bar{y}'  .
\end{equation}
To perform this integral, one must  invert the transformations given in Section 3 to give $\bar{x}$ and $\bar{y}$ in terms of the initial variables $S$ and $f$. In fact, one has  that
\begin{equation}
\left\{ \begin{array}{l}
\bar{x}=\frac{1}{2}\left(\frac{\ln S -\left(r-\frac{1}{2} \sigma^{2}\right)(T-\tau)}{\sigma}+\frac{f}{\Gamma}\right)-\frac{\mu_{f}(T-\tau)}{2 \Gamma} \\
\bar{y}=\frac{1}{2}\left(\frac{\ln S-\left(r-\frac{1}{2} \sigma^{2}\right)(T-\tau)}{\sigma}-\frac{f}{\Gamma}\right)+\frac{\mu_{f}(T-\tau)}{2 \Gamma} ,
\end{array} \right.
\end{equation}
so
\begin{equation}
\bar{x}-\bar{x}'=\frac{1}{2 \sigma} \ln \left(\frac{S}{S'}\right)+\frac{1}{2 \Gamma}(f-f') ,
\end{equation}
and
\begin{equation}
\bar{y}-\bar{y}'=\frac{1}{2 \sigma} \ln \left(\frac{S}{S'}\right)-\frac{1}{2 \Gamma}\left(f-f'\right) .
\end{equation}
Also
\begin{equation}
d \bar{x} d \bar{y}=\frac{1}{2 \sigma \Gamma S} d S d f ,
\end{equation}
so equation (\ref{solution_generic_gauss}) becomes
\begin{equation}
\begin{aligned}
\psi(S, f, \tau)=\int_{0}^{+\infty}  \int_{-\infty}^{+\infty} \frac{e^{-\left[\frac{1}{2 \sigma} \ln \left(\frac{S}{S'}\right)+\frac{1}{2 \Gamma}\left(f-f'\right)+\alpha_{\bar{x}} \tau \right]^{2}}}{\sqrt{2 \pi \tau}} \times \\
\delta\left[\frac{1}{2 \sigma} \ln \left(\frac{S}{S'}\right)-\frac{1}{2 \Gamma}\left(f-f^{\prime}\right)+\alpha_{\bar{y}} \tau \right] \Phi\left(S', f'\right) \frac{d S' d f'}{2 \sigma \Gamma S'} .
\end{aligned}
\end{equation}
After performing the $f'$ integral, one gives
\begin{equation} \label{gaussian_psi}
\psi(S, f, \tau)=\int_{0}^{+\infty} \frac{e^{-\frac{\left[\ln \left(\frac{S}{S'}\right) +\left(\alpha_{\bar{x}}+\alpha_{\bar{y}} \right) \sigma \tau \right]^{2}}{2 \sigma^{2} \tau}}}{\sqrt{2 \pi \sigma^{2} \tau}}  \Phi\left(S', f_0 \right)  \frac{d S'}{S'} ,
\end{equation}
where
\begin{equation} \label{gaussian_f0}
f_{0}=f_0(S,S',f, \tau)= f-\frac{\Gamma}{\sigma} \ln \left(S / S'\right)-2 \Gamma \alpha_{\bar{y}} \tau .
\end{equation} \\
Two obtain an explicit analytic solution, one can consider now the case of a pure Call, for which the contract function $\Phi$ is
\begin{equation} \label{purecall}
\Phi\left(S , f \right) = \Phi\left(S \right) = max\{0, S-K \} ,
\end{equation}
so
\begin{equation}
\psi(S, f, \tau)= \int_{K}^{+\infty} \frac{e^{-\frac{\left[\ln \left(\frac{S}{S'}\right) +\left(\alpha_{\bar{x}}+\alpha_{\bar{y}} \right) \sigma \tau \right]^{2}}{2 \sigma^{2} \tau}}}{\sqrt{2 \pi \sigma^{2} \tau}}  (S' - K) \frac{d S'}{S'} .
\end{equation} 
The last integral can be performed exactly to give
\begin{equation} \label{BSformulageneric_1}
\psi(S, f, \tau)= e^{\sigma\left(\alpha_{\bar{x}}+\alpha_{\bar{y}}\right) \tau+\frac{1}{2} \sigma^{2} \tau} \ S \ N\left(d_{1}\right)-E \  N\left(d_{2}\right) ,
\end{equation}
where
\begin{equation}  \label{BSformulageneric_2}
d_{1}=\frac{\ln (S/E)+\sigma\left(\alpha_{\bar{x}}+\alpha_{\bar{y}} \right) \tau+\sigma^{2} \tau}{\sigma \sqrt{\tau}} ,
\end{equation}
and
\begin{equation}  \label{BSformulageneric_3}
d_{2}=\frac{\ln (S/E) + \sigma \left(\alpha_{\bar{x}}+ \alpha_{\bar{y}}\right) \tau}{\sigma \sqrt{\tau}} .
\end{equation} 
\subsubsection{The solutions for weak limit of the Gaussian bubble}
For the weak limit of the Gaussian model (\ref{BSE_xbar_ybar_f=0_gauss}), one has 
\begin{equation}
\begin{aligned}
\alpha_{\bar{x}} = & \  \frac{(r - \mu)}{2 \sigma}  , \\
\alpha_{\bar{y}} = & -\frac{(r - \mu)}{2 \sigma} ,
\end{aligned}
\end{equation}
The option price given by (\ref{Vpsi_gauss}) is 
\begin{equation}
V(S, f, \tau)= e^{-r\tau} \psi(S, f, \tau) .
\end{equation}
Then, due that 
\begin{equation}
\alpha_{\bar{x}}+\alpha_{\bar{y}}=0 ,
\end{equation}
by using (\ref{BSformulageneric_1}), (\ref{BSformulageneric_2}) and (\ref{BSformulageneric_3}), one finds that the option price in the weak limit of the Gaussian model is
\begin{equation} 
V(s, f, \tau)=e^{-r \tau} \cdot\left[e^{\frac{1}{2} \sigma^{2} \tau} \ S \ N\left(d_{1}\right)-E \  N\left(d_{2}\right)\right] ,
\end{equation}
or
\begin{equation} \label{optionpricegaussian_weak_1}
V(s, f, \tau)= e^{-(r-\frac{1}{2} \sigma^{2}) \tau} \ S \ N\left(d_{1}\right)-E e^{-r \tau} \  N\left(d_{2}\right) ,
\end{equation}
with
\begin{equation}  \label{optionpricegaussian_weak_2}
d_{1}=\frac{\ln (S/E)+\sigma^{2} \tau}{\sigma \sqrt{\tau}} ,
\end{equation}
and
\begin{equation}  \label{optionpricegaussian_weak_3}
d_{2}=\frac{\ln (S/E) }{\sigma \sqrt{\tau}} .
\end{equation}
\subsubsection{The solutions for strong limit of the Gaussian bubble}
For the strong limit of the Gaussian model (\ref{BSE_xbar_ybar_f=infinito_gauss}) one has 
\begin{equation}
\begin{aligned}
\alpha_{\bar{x}} = & -\frac{(r - \mu)}{2 \sigma} ,  \\
\alpha_{\bar{y}} = & -\frac{(r - \mu)}{2 \sigma} ,
\end{aligned}
\end{equation}
then
\begin{equation}
\alpha_{\bar{x}}+\alpha_{\bar{y}}= -\frac{(r - \mu)}{\sigma} ,
\end{equation}
so by (\ref{BSformulageneric_1}), (\ref{BSformulageneric_2}) and (\ref{BSformulageneric_3}) the function $\Psi$ is
\begin{equation}
\Psi(S, f, \tau)= e^{-(r - \mu) \tau+\frac{1}{2} \sigma^{2} \tau} \ S \ N\left(d_{1}\right)-E \  N\left(d_{2}\right) ,
\end{equation}
with
\begin{equation} \label{optionpricegaussian_strong_2}
d_{1}=\frac{\ln (S/E)+(r - \mu) \tau+\sigma^{2} \tau}{\sigma \sqrt{\tau}} ,
\end{equation}
and
\begin{equation} \label{optionpricegaussian_strong_3}
d_{2}=\frac{\ln (S/E) + (r - \mu) \tau}{\sigma \sqrt{\tau}} .
\end{equation}
The option price is given in this case by (\ref{Vpsi_gauss_2}) 
\begin{equation}
V(S, f, \tau)= e^{-\mu \tau} \Psi(S, f, \tau) ,
\end{equation}
so the option price in the strong limit of the Gaussian bubble is
\begin{equation}
V(S, f, \tau)=e^{- \mu \tau} \left[e^{-(r - \mu) \tau+\frac{1}{2} \sigma^{2} \tau} \ S \ N\left(d_{1}\right)-E \  N\left(d_{2}\right)\right] ,
\end{equation}
or
\begin{equation} \label{optionpricegaussian_strong_1}
V(S, f, \tau)= e^{-(r-\frac{1}{2} \sigma^{2})\tau} \ S \ N\left(d_{1}\right)-E e^{- \mu \tau} \  N\left(d_{2}\right) .
\end{equation} \\ 

\subsubsection{The solutions for the $f/\sigma \approx -1$ case for the Gaussian bubble}
For the case $f/\sigma  \approx -1$, the dynamics are given by equation (\ref{BSE_xbar_ybar_gauss_f=-1}), which is a special case of (\ref{solution_generic_xbar_ybar}),  with
\begin{equation}
\begin{array}{l}
\alpha_{\bar{x}}=0 ,\\
\alpha_{\bar{y}}= -\frac{(r-\mu)}{2 \sigma} ,
\end{array}
\end{equation}
The solution is given then according to (\ref{BSformulageneric_1}),  (\ref{BSformulageneric_1}), (\ref{BSformulageneric_3}) and (\ref{Vpsi_gauss_f=-1}) by
\begin{equation} \label{solution_V_f=-1}
V=e^{-\left(r-\frac{1}{2} \sigma^{2}\right) \tau} \ S \ N\left(d_{1}\right)-E \ e^{-\frac{\left(r+\mu\right)}{2} \tau} \ N \left(d_{2}\right) ,
\end{equation}
with
\begin{equation}
d_{1}=   \frac{\ln (S/E)-\frac{(r-\mu)}{2} \tau +\sigma^{2} \tau}{\sigma \sqrt{\tau}} ,
\end{equation}
and
\begin{equation}
d_{2}=   \frac{\ln (S/E)-\frac{(r-\mu)}{2} \tau}{\sigma \sqrt{\tau}} .
\end{equation} 
 
\subsection{The solutions for the lognormal bubble}

The weak and  strong limits of the lognormal bubble are given by equations (\ref{BSE_xbar_ybar_f=0_lognormal}), (\ref{BSE_xbar_ybar_f=infinito_lognormal}), which are again of the form of equation (\ref{BSEffreegeneralcase}), so the solution in the $(x,y,\tau)$ is given by (\ref{solution_generic}). Now one can map this solution into the $(S, f, \tau)$ space by taking the inverse of the transformation done in Section 4. The result is
\begin{equation} \label{x_todas_las_otras}
x=\frac{1}{2}\left(\frac{\ln S-\left(r-\frac{1}{2} \sigma^{2}\right)(T-\tau)}{\sigma}+\frac{\ln f-\left(\bar{u}_{f}-\frac{1}{2} \bar{\Gamma}^{2}\right)(T-\tau)}{\bar{\Gamma}} \right) ,
\end{equation}
\begin{equation} \label{y_todas_las_otras}
y=\frac{1}{2}\left(\frac{\ln S-\left(r-\frac{1}{2} \sigma^{2}\right)(T-\tau)}{\sigma}-\frac{\ln f-\left(\bar{u}_{f}-\frac{1}{2} \bar{\Gamma}^{2}\right)(T-\tau)}{\bar{\Gamma}} \right) ,
\end{equation}
so
\begin{equation}
\begin{array}{l}
x-x'=\ln \left[\left(\frac{S}{S'}\right)^{1 / 2 \sigma} \ \left(\frac{f}{f'}\right)^{1 / 2 \bar{\Gamma}}\right] , \\
\end{array}
\end{equation}
and
\begin{equation}
\begin{array}{l}
y-y'=\ln \left[\left(\frac{S}{S'}\right)^{1 / 2 \sigma} \ \left(\frac{f}{f'}\right)^{-1 / 2 \bar{\Gamma}}\right] .\\
\end{array}
\end{equation}
Also, one can show that 
\begin{equation}
dx dy = \frac{1}{2 \sigma \bar{\Gamma} S f} \ dS \ df .
\end{equation}
In this way, the solution in the $(S, f, \tau)$ space is then
\begin{equation} \label{solutiongeneric_2}
\begin{aligned}
\psi(S, f, \tau)= & \int_{-\infty}^{\infty} \int_{-\infty}^{\infty}  \frac{1}{\sqrt{2 \pi \tau}} e^{-\frac{\left(\ln \left[\left(\frac{S}{S'}\right)^{1 / 2 \sigma} \ \left(\frac{f}{f'}\right)^{1 / 2 \bar{\Gamma}}\right]+ \alpha_{x} \tau \right)^{2}}{2 \tau}} \times \\
&\delta \left(\ln \left[\left(\frac{S}{S'}\right)^{1 / 2 \sigma} \ \left(\frac{f}{f'}\right)^{-1 / 2 \bar{\Gamma}}\right]+\alpha_{y} \tau \right) \Phi\left(S', f' \right) \  \frac{1}{2 \sigma \bar{\Gamma} S' f'} \ dS' \ df' .
\end{aligned}
\end{equation}
By integrating in $f'$, one obtains 
\begin{equation} \label{lognormal_psi}
\psi(S, f, \tau)= \int_{0}^{\infty}  \frac{1}{\sqrt{2 \pi \sigma^2 \tau}} e^{-\frac{\left(\ln \left[\left(\frac{S}{S'}\right)^{1 / \sigma} \right] + (\alpha_{x} + \alpha_{y}) \tau \right)^{2}}{2 \tau}} \ \Phi\left(S', f_{0} \right) \  \frac{dS'}{ S'}  ,
\end{equation}
where $f_0$ denotes on this occasion the function
\begin{equation} \label{lognormal_f0}
f_{0} = f_{0}(S,S',f) = f \left(\frac{S^{\prime}}{S}\right)^{\bar{\Gamma} / \sigma} e^{-2 \bar{\Gamma} \alpha_y \tau} .
\end{equation}
Note that this is the same result obtained in (\ref{gaussian_psi}), but  the form of $f_0$ is different. \\ \\
Thus, if one considers a pure Call contract as in (\ref{purecall}),  then (\ref{lognormal_psi}) implies that the generic solution for the pure Call contract is given by equations (\ref{BSformulageneric_1}),  (\ref{BSformulageneric_2}) and (\ref{BSformulageneric_3}) but with $\alpha_{\bar{x}}$ and $\alpha_{\bar{y}}$ replaced by $\alpha_x$ and $\alpha_y$, respectively.
\subsubsection{The solutions for weak limit of the lognormal bubble}
For the weak limit, equation (\ref{BSE_xbar_ybar_f=0_lognormal})
implies that 
\begin{equation}
\begin{aligned}
\alpha_{x} = & \  \frac{(r - \mu)}{2 \sigma}  , \\
\alpha_{y} = & -\frac{(r - \mu)}{2 \sigma}  ,
\end{aligned}
\end{equation}
so
\begin{equation}
\alpha_{x}+\alpha_{y}=0 ,
\end{equation}
and the solutions for the option price $V$ are given again by equations (\ref{optionpricegaussian_weak_1}), (\ref{optionpricegaussian_weak_2}) and (\ref{optionpricegaussian_weak_3}).

\subsubsection{The solutions for strong limit of the lognormal bubble}
For the strong limit, equation (\ref{BSE_xbar_ybar_f=infinito_lognormal})
implies that 
\begin{equation}
\begin{aligned}
\alpha_{x} = & -\frac{(r - \mu)}{2 \sigma} ,  \\
\alpha_{y} = & -\frac{(r - \mu)}{2 \sigma} ,
\end{aligned}
\end{equation}
so
\begin{equation}
\alpha_{x}+\alpha_{y}=-\frac{(r - \mu)}{\sigma} ,
\end{equation}
and the solutions for the option price $V$ are given this time   by equations (\ref{optionpricegaussian_strong_1}), (\ref{optionpricegaussian_strong_2}) and (\ref{optionpricegaussian_strong_3}). \\ \\

\begin{figure}[h!] \label{figure1}
	\centering
	\includegraphics[scale=0.5]{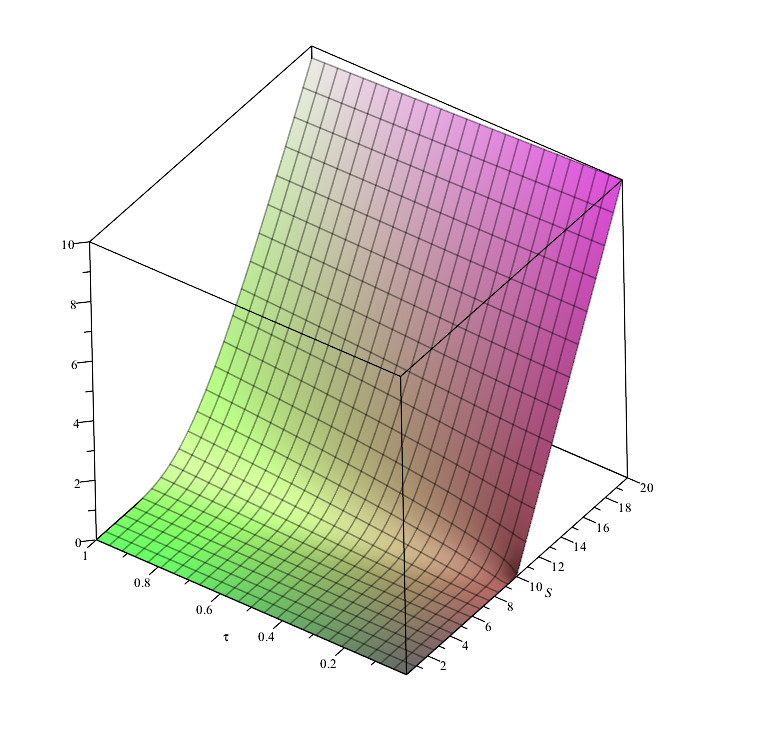}
	\includegraphics[scale=0.5]{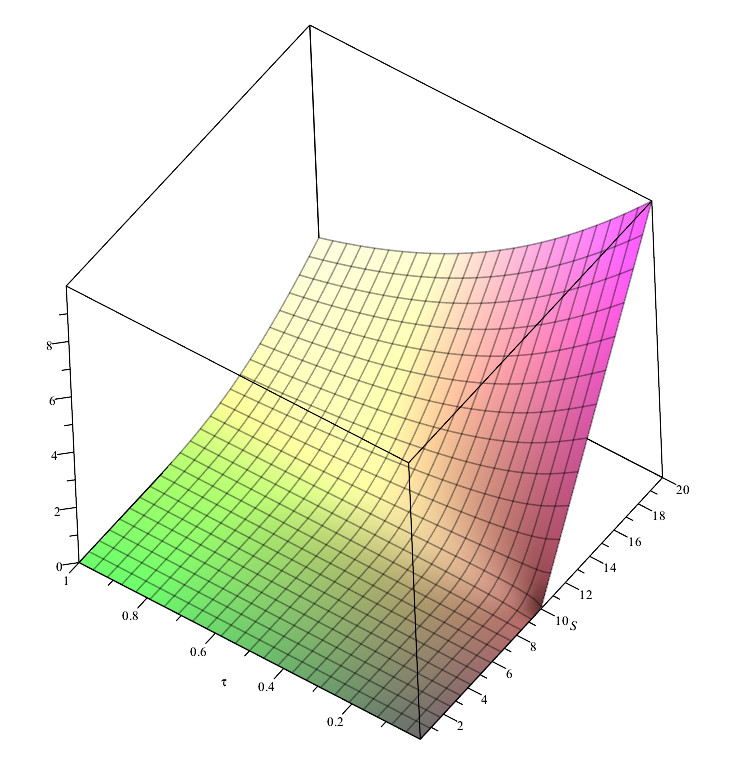} 
	\includegraphics[scale=0.5]{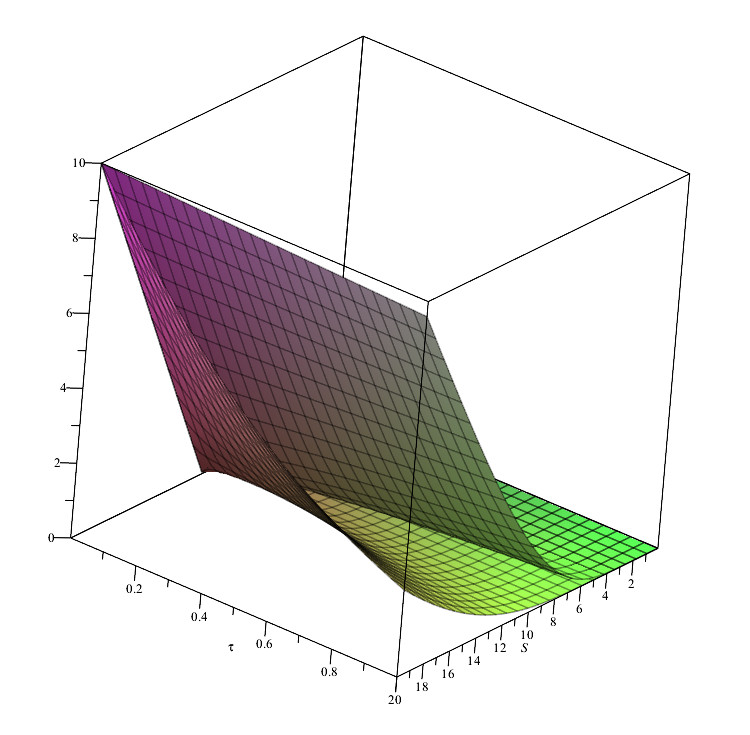}
	\caption{From left to right: weak solution, strong solution  and both solutions for $E=10$, $\mu = 0.8$, $r=0.2$, $\sigma=0.4$ in the pure Call case.}
\end{figure}

Figures (\ref{figure1})  and (\ref{figure1}) show the behavior of the weak and strong solution $V$ for two different parameter sets.
\noindent \\
\begin{figure}[h!] \label{figure2}
	\centering
	\includegraphics[scale=0.5]{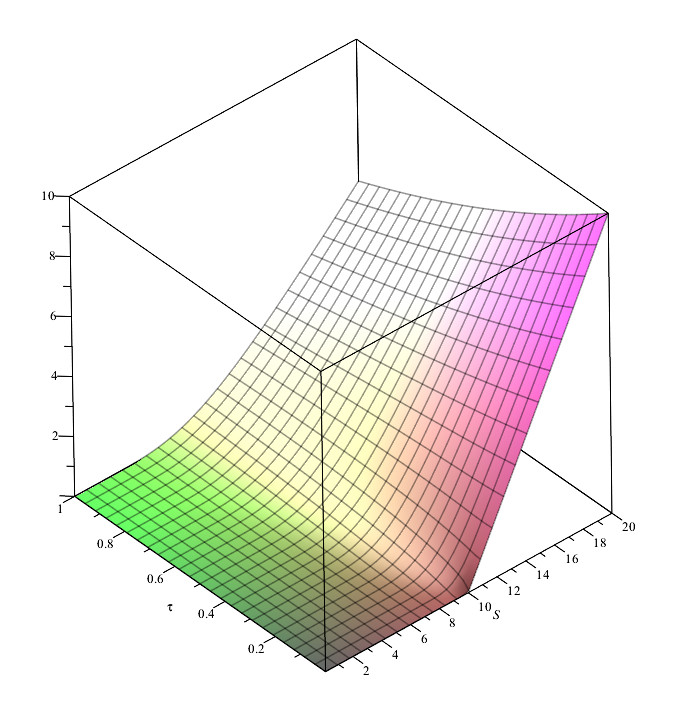}
	\includegraphics[scale=0.5]{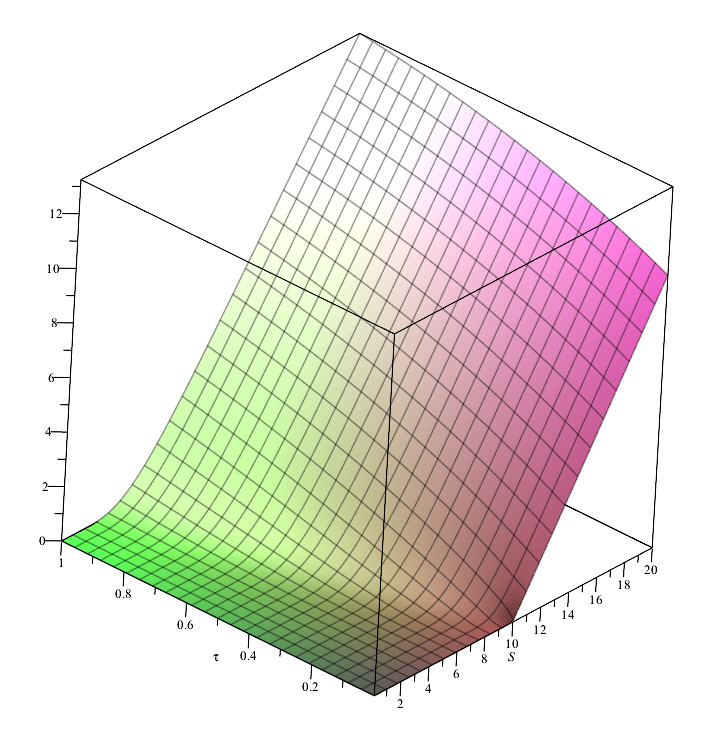} 
	\includegraphics[scale=0.5]{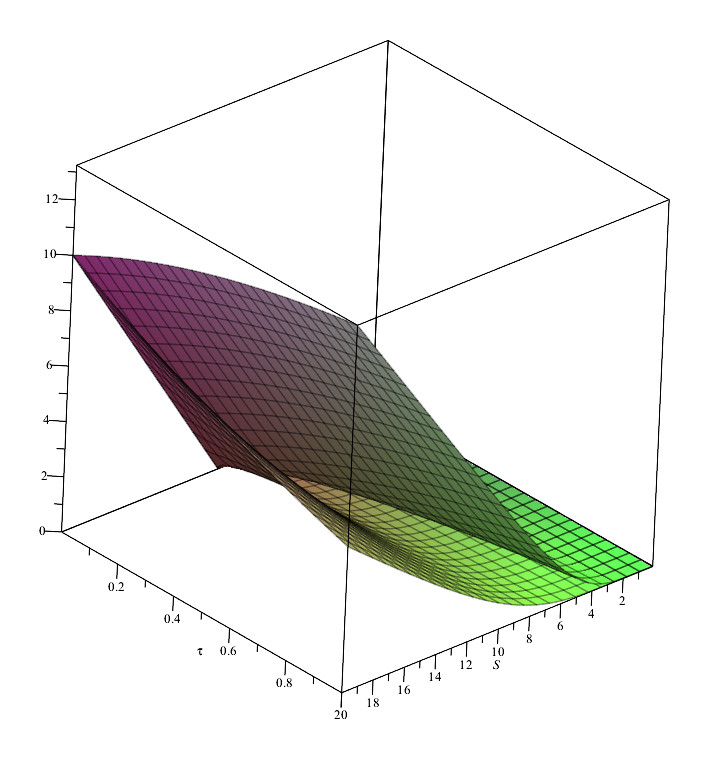}
	\caption{From left to right: weak solution, strong solution  and both solutions for  $E=10$, $\mu = 0.2$, $r=0.8$, $\sigma=0.4$ in the pure Call case.}
\end{figure}

Finally, one must  note that all of these results are valid if $\mu$, $\sigma$, $\mu_f$ and $\Gamma$ are functions of the space--time variables $(S,f, t)$ that satisfy the asymptotic behavior
\begin{equation}
\begin{aligned}
\lim_{f \rightarrow 0}  &\ \mu(S,f, t) \  &\approx&\ \  \mu^0  \\
\lim_{f \rightarrow 0}  &\ \sigma(S,f, t) \ &\approx&\ \ \sigma^0  \\
\lim_{f \rightarrow 0}  &\ \mu_{f}(S,f, t) \  &\approx&\ \ \mu_{f}^0  \\
\lim_{f \rightarrow 0}  &\ \Gamma(S,f, t) \ &\approx&\ \ \Gamma^0
\end{aligned} \ \ \ \ \ \ \ \ \ \ \  \ \ \ \ \
\begin{aligned}
\lim_{f \rightarrow \infty}  &\ \mu(S,f, t) \  &\approx&\ \ \mu^{\infty}  \\
\lim_{f \rightarrow \infty}  &\ \sigma(S,f, t) \ &\approx&\ \  \sigma^{\infty}  \\
\lim_{f \rightarrow \infty}  &\ \mu_{f}(S,f, t) \  &\approx&\ \ \mu_{f}^{\infty}  \\
\lim_{f \rightarrow \infty}  &\ \Gamma(S,f, t) \ &\approx&\ \ \Gamma^{\infty} ,
\end{aligned}
\end{equation}
for the Gaussian Bubble or \\
\begin{equation}
\begin{aligned}
\lim_{f \rightarrow 0}  &\ \mu(S,f, t) \  &\approx&\ \  \mu^0  \\
\lim_{f \rightarrow 0}  &\ \sigma(S,f, t) \ &\approx&\ \ \sigma^0  \\
\lim_{f \rightarrow 0}  &\ \mu_{f}(S,f, t) \  &\approx&\ \ f \  \mu_{f}^0  \\
\lim_{f \rightarrow 0}  &\ \Gamma(S,f, t) \ &\approx&\ \ f \ \Gamma^0
\end{aligned} \ \ \ \ \ \ \ \ \ \ \   \ \ \ \ \
\begin{aligned}
\lim_{f \rightarrow \infty}  &\ \mu(S,f, t) \  &\approx&\ \ \mu^{\infty}  \\
\lim_{f \rightarrow \infty}  &\ \sigma(S,f, t) \ &\approx&\ \  \sigma^{\infty}  \\
\lim_{f \rightarrow \infty}  &\ \mu_{f}(S,f, t) \  &\approx&\ \ f \ \mu_{f}^{\infty}  \\
\lim_{f \rightarrow \infty}  &\ \Gamma(S,f, t) \ &\approx&\ \ f \ \Gamma^{\infty} ,
\end{aligned}
\end{equation}
for the log-normal bubble. Here, $\mu^0$, $\sigma^0$, $\mu_{f}^0$, $\Gamma^0$, $\mu^\infty$, $\sigma^\infty$, $\mu_{f}^\infty$ and $\Gamma^\infty$ are constant. \\

\section{Conclusions}
In this article, a stochastic model of endogenous arbitrage bubbles was developed. In this case, the arbitrage bubble satisfies a stochastic differential equation (\ref{df}), and the option price is given by the general equation (\ref{BSf_equation}). This equation has several interesting limit behaviors. For example, for $\Gamma = 0$ in (\ref{BSf_equation}), there exist both ``weak'' $f \approx 0$  and ``strong'' $f \rightarrow \infty$ bubble regimens. The weak case corresponds to the usual arbitrage-free Black--Scholes model, while the strong case also corresponds to a Black--Scholes model where the interest rate has been changed by the mean value of the underlying assets. \\ \\
For the case $\Gamma \ne 0$, it has been shown that similar weak and strong bubble behaviors exist for two different stochastic bubbles: the Gaussian and the lognormal bubbles. For a pure Call contract case, the dynamic equations of these weak and stronger limits are given by equations (\ref{BSE_xbar_ybar_f=0_gauss}), (\ref{BSE_xbar_ybar_f=infinito_gauss}) and (\ref{BSE_xbar_ybar_f=0_lognormal}) and (\ref{BSE_xbar_ybar_f=infinito_lognormal}), respectively.
The solutions of these asymptotic equations are given by equations (\ref{optionpricegaussian_weak_1}) and (\ref{optionpricegaussian_strong_1}), which are equivalents to the Black--Scholes solution but with $\Gamma \ne 0$. \\
It is interesting to note that for the Gaussian bubble case, where $f$ can take positive and negative values, there exist another weak limit $f/\sigma \approx -1$, whose dynamics are given by (\ref{BSE_xbar_ybar_gauss_f=-1}) with a solution given by (\ref{solution_V_f=-1}). However, for the lognormal case, that limit cannot be reached because $f$ would always maintain positive according to (\ref{fpositive_lognornal}). \\ \\
Thus, the usual Black--Scholes theory can be considered as only an asymptotic limit of a more general model given by equation (\ref{BSf_equation}). Although the solutions studied here are limit cases of the general model (\ref{BSf_equation}),  they are by no means important. Furthermore, these solutions can test the accuracy of the general case's numerical solution in the different asymptotic scenarios. \\ \\
In a forthcoming article, I will obtain the corresponding numerical solutions of (\ref{BSf_equation}) and compare them with the weak and strong limits solutions obtained in this paper.

\end{document}